# Copula Component Analysis

Jian Ma and Zengqi Sun

Department of Computer Science, Tsinghua University, Beijing, China, 100084

majian03@mails.tsinghua.edu.cn

(Dated: March 20, 2007)

*Abstract*

A framework named Copula Component Analysis for blind source separation is proposed as a generalization of ICA. It differs from ICA which assumes independence of sources that the underlying components may be dependent with certain structure which is represented by Copula. By incorporating dependency structure, much accurate estimation can be made in principle in the case that the assumption of independence is invalidated. A two phrase inference method is introduced for CCA which is based on the notion of multidimensional ICA.

## 1 Introduction

Blind Source Separation (BSS) is to recover the underlying component from their mixtures, where the mixing matrix and distribution of component are unknown. To solve this problem, Independent component analysis (ICA) is the most popular method to extract those components under the assumption of statistically independence.[1]-[5] However, in practice, the independent assumption of ICA model can not be satisfied and thus strongly confines its application. Many works have been contributed to extend the ICA model by relaxing the independent assumption,[6] such as Tree-ICA[7], Topology ICA[8]. The central problem is how to break the independent assumption and to incorporate dependency structure for different kinds of problem into the model.

Copula [9] is a recently developed mathematical theory for multivariate probability analysis. It separates joint probability distribution function into the product of marginal distributions and a Copula function which represents the dependency structure of random variables. According to *Sklar* theorem, given a joint distribution with margins, there exists a copula uniquely determined. Through Copula, we can clearly represent the dependent relation of variables and analysis multivariate distribution of the underlying components.

The aim of this paper is using Copula to model the true distribution with dependence structure of the underlying components. By doing this, we transform BSS into a parametric or semi-parametric estimation problem which mainly concentrate on the estimation of dependency structure while identifying the independent components as ICA do.

This paper is organized as follows: for self contained, we briefly review ICA and its extensions in section 2. The main conclusions on Copula are introduced in section 3. In section 4, we propose a new model for BSS, named *Copula Component Analysis* (CCA). Inference method for CCA is presented in section 5. Finally in section 6 we conclude the paper and give some further research directions.

## 2 ICA and its Extensions

Given a random vectors **x**, ICA is modeled as

$$\mathbf{x} = \mathbf{As} \quad \text{or} \quad \hat{\mathbf{s}} = \mathbf{Wx}, \tag{1}$$

where the source signal $\mathbf{s} = [s_1, \cdots, s_n]$ assume to be mutually independent, $\mathbf{A}$ and $\mathbf{W} = \mathbf{A}^-$ is the invertible mixing and demixing matrix to be solved so that the recovered underlying components $\hat{\mathbf{s}}$ is as statistically independent as possible.

Statistical independence of sources means that the joint probability density of $\mathbf{x}$ and $\mathbf{u}$ can be factorized as

$$p(\mathbf{x}) = p(\mathbf{As}) = |\det(\mathbf{W})| p(\mathbf{s})$$
$$p(\mathbf{s}) = \prod_{i=1}^{n} p_i(s_i)$$
(2)

The community has presented many extensions of ICA with different types of dependency structures, such as Tree-ICA and Topographical ICA. Bach [7] assumes that dependency can be modeled as a tree (or a forest). After the contrast function was extended with T-mutual information, Tree-ICA tries to find a mixing matrix A and tree structure T simultaneously by embedding a Chow-Liu algorithm into algorithm. Hyvärinen [8] introduce the variance into ICA model so as to model dependency structure.

Cardoso generalized the notion of ICA into multidimensional ICA.[6] In such a model, the components are grouped into many clusters whose joint density can be presented as

$$p(\mathbf{u}) = \prod_{k=1}^{m} p_k(\mathbf{u}_k).$$
(3)

where $\mathbf{u}_k = (u_{i_k}, \ldots, u_{i_{k+1}-1})$ is a random vector, and $p_k(\mathbf{u}_k) = p_k(u_{i_k}, \ldots, u_{i_{k+1}-1})$ is the joint pdf of $\mathbf{u}_k$, which can not be factorized into the product of their marginal pdf.

## 3. Copula

3.1 Introduction

Copula is a recently introduced theory which separates the margin law and the joint law and therefore gives dependency structure as a function. According to Nelson [9], it defines as follows:

**Definition (Copula)**: A bidimensional copula is a function $C(x, y): R^2 \mapsto R$ with following properties:

1) $(x, y) \subset I^2$;

2) $C(x_2, y_2) - C(x_1, y_2) - C(x_2, y_1) + C(x_1, y_1) \geq 0$, for $x_1 \leq x_2$ and $y_1 \leq y_2$;

3) $C(x, 1) = x$ and $C(1, y) = y$.

It's not hard to know that such defined $C(x, y)$ is a cdf on $I^2$. Multidimensional Copula can be generalized in a same manner which presents in [9].

**Theorem 1** *Sklar Theorem*. Given a multidimensional variable $\mathbf{x} = (x_1, \ldots, x_n) \in R^n$ with their corresponding cumulant function and density function $u_i = F_i(x_i)$ and $p_i(x_i), i = 1,\ldots,n$. Let $F(\mathbf{x}): R^n \mapsto I$ denotes the joint distribution, then there exists a Copula $C(\bullet): I^n \mapsto I$ so that

$$F(\mathbf{x}) = C(u_1,\ldots,u_n). \tag{4}$$

If the copula is differentiable, the joint density function of $F(\mathbf{x})$ is

$$p_{1,\ldots,n}(\mathbf{x}) = \prod_{i=1}^{n} p_i(x_i) C'(\mathbf{u}). \tag{5}$$

where $\mathbf{u} = (u_1,\ldots,u_n)$ and $C'(\mathbf{u}) = \dfrac{\partial^n C(\mathbf{u})}{\partial u_1,\ldots,\partial u_n}$.

Given a random vector $\mathbf{x} = (x_1,\ldots,x_n)$ with mutually independent variables, and their cdf $F(\mathbf{x}) = \prod_i F_i(x_i)$. It is easy to obtain that the corresponding copula function called *Product Copula* is $C(\mathbf{u}) = \prod_i u_i$ and $C'(\mathbf{u}) = 1$.

3.2 Classes of Copula

*Gaussian copula* is a popular and easy-to-use one because of its easily computation and clearly meaningful parameters:

$$C(x;\rho) = |\rho|^{-\frac{1}{2}} \exp\left(-\frac{1}{2}\left(x^T \left(\rho^{-1} - I\right) x\right)\right), \tag{18}$$

where $\rho$ is correlation matrix of random vector $x$.

Archimedean copula is a very important and widely used class of copulas due to its ease of construction, variety of families and nice property it possessed. A multidimensional Archimedean copula is defined as

$$C(x;\theta) = \Phi^{-1}\left(\sum_{i=1}^{n} \Phi(x_i;\theta);\theta\right), \tag{19}$$

where the invertible function $\Phi(\bullet;\theta)$ with parameter vector $\theta$ is called generator.

## 4. Copula Component Analysis

4.1 Geometry of CCA
As privously stated, ICA assumes that the underlying components are mutually independent, which can be represented as (2). CCA also use the same representation (2) as ICA, but without independence assumption. Here, Let the joint density function represents by Copula:

$$p_c(\mathbf{x}) = \prod_{i=1}^{N} p_i(x_i) C'(\mathbf{u}). \tag{6}$$

where the denpendency structure is modeled by function $C(\mathbf{u})$.

The goal of estimation is to minimize the distance between the real pdf of random vector $\mathbf{x}$ and its counterpart of the proposed model. Given a random vector $\mathbf{x}$, with pdf $p(\mathbf{x})$, the distance between $p(\mathbf{x})$ and $p_c(\mathbf{x})$ in a sense of *K-L* divergence can be represents as

$$D(p \| p_c) = E_{p(\mathbf{x})} \log \frac{p(\mathbf{x})}{p_c(\mathbf{x})} = E_{p(\mathbf{x})} \log \frac{p(\mathbf{x})}{\prod_i p_i(x_i)} - E_{p(\mathbf{x})} \log C'(\mathbf{u}). \tag{7}$$

The first term of (7) is corresponding to the K-L divergence between $p(\mathbf{x})$ and ICA model and the second term is corresponding to entropy of Copula $C(\mathbf{x})$.

**Theorem 2** Given a random vector $\mathbf{x} = (x_1, \ldots, x_n) \in R^n$ with pdf $p(\mathbf{x})$ and its joint pdf $p_c(\mathbf{x}) = \prod_{i=1}^{n} p_i(x_i) C'(\mathbf{u})$, where $u_i = F_i(x_i)$ is the cdf of $x_i$ and dependency structure is presented by Copula function $C(\mathbf{u}): I^n \mapsto I, \mathbf{u} \in R^n$ and $C'(\mathbf{u}) = \frac{\partial^n C(\mathbf{u})}{\partial u_1, \ldots, \partial u_n}$. The K-L divergence $D(p \| p_c)$ is as

$$D(p \| p_c) = I(x_1, \ldots, x_n) + H(C'(\mathbf{u})). \tag{8}$$

where $H(\cdot)$ is the Shannon differential entropy. That is, the K-L divergence between $p(\mathbf{x})$ and $p_c(\mathbf{x})$ equal to the difference of the mutual information between $\mathbf{x} = (x_1, \ldots, x_n)$ and entropy H for function $u \sim C'$.

Using the invariant of K-L divergence, we now have the following corollary to theorem 2 for BSS problem $\mathbf{s} = \mathbf{W}\mathbf{x}$:

*Corollary 1* With the same denotation of Theorem 2, the K-L divergence for BSS problem is

$$D(p \| p_c) = I(s_1, \ldots, s_n) + H(C'(\mathbf{u}_s)), \tag{9}$$

where $\mathbf{u}_s$ denotes the marginal variable for $s$ and we assume the number of source equals to that of sensors.

In other words, the distance between ICA model and the true model is presented by dependency structure and its value equals to entropy of the underlying Copula function. It can be easily learned from (8) that incorporating the dependency structure into model, the distance between data and model can be further closer than that of ICA model.

ICA is a special case when it assumes mutual independence of underlying components. Actually, ICA only minimizes the first part of (8) under the assumption of independence. This also explains why sometime ICA model is not applicable when dependency exists.

## 4.2 Multidimensional ICA

From the notion of multidimensional ICA generalized from ICA by Cardoso, it can be derived that

$$p(\mathbf{x}) = \prod_{k=1}^{m} p_k(\mathbf{x_k}) = \prod_{k=1}^{m} p_k\left(x_{i_k}, \ldots, x_{i_{k+1}-1}\right) = \prod_{k=1}^{m} \prod_{l=i_k}^{i_{k+1}-1} p_k(x_l) C_k'(\mathbf{u_k}) \\ = \prod_{i=1}^{n} p_i(x_i) \prod_{k=1}^{m} C_k'(\mathbf{u_k}). \tag{10}$$

Where $C_k(\bullet)$ is the copula with respect to $p_k(\bullet)$. On the other side, the definition of Copula tells

$$p(\mathbf{u}) = \prod_{i=1}^{n} p_i(u_i) C'(\mathbf{u}). \tag{11}$$

According to *Sklar theorem*, if all $p_i(\bullet)$ exists, then $C$ is unique. Thus we can derive the following result:

**Theorem 3** The copula corresponding to Multidimensional ICA is *factorial* if all the marginal pdf of component exists, that is

$$C'(\mathbf{u}) = \prod_{k=1}^{m} C_k'(\mathbf{u_k}). \tag{12}$$

Proof. Because of the unique of *C*, the above (12) can be easily derived by comparing (10) and (11). The theorem can guide hypothesis selection of Copula. Copula could be factorized as a product of sub-function with different type for dependency structure of different sub-space.

Combined (8) and (12), we can derived the following

$$D(p \| p_c) = I(u_1, \ldots, u_k) + \sum_{k=1}^{m} H(C_k'). \tag{13}$$

It means that the distance between the true model and ICA model composes of entropy of Copulas which corresponds to every un-factorial ICs spaces. Therefore, if we want to get the true model as much closer than ICA, we can find dependency structure of each space, that is, approach the goal step by step. This is the guide principle for design of algorithm of copula component analysis.

## 5 Inference of CCA

### 5.1 General Framework

We study inference method of CCA based on the notion of multidimensional ICA. Suppose the underlying copula function parameterized by $\theta \in \Theta$, thus the estimation of CCA should infer the demixing matrix $\mathbf{W}$ and $\theta$. According to theorem 2, estimation of the underlying sources through

our model requires the minimization of the K-L divergence of (8) or (13). Thus the objective function is

$$\min D(p \| p_c; \mathbf{W}, \theta), \qquad (14)$$

which composes of two sub-objective: $\min I(x_1, \ldots, x_n; \mathbf{W})$ and $\max H(C'(\mathbf{u}); \mathbf{W}, \theta)$. Because **u** in the latter objective depends on the structure of IC spaces derived from the former objective, we should minimize $\min I(x_1, \ldots, x_n; \mathbf{W})$ first. The first objective can be done by ICA algorithm and for the second one we proposed the Infomax principle given a parametric family of copula.

We propose that the framework of CCA composes of two phrases:

1) Solve **W** through minimization of mutual information $I(x_1, \ldots, x_n; \mathbf{W})$.

2) Determine $\theta$ in $C(\mathbf{u}; \theta)$ so that entropy of copula is maximized.

5.2 Maximum Likelihood Estimation

Given the parametric model of Copula, maximum likelihood estimation can be deployed under the constraint of ICA. Consider a group of independent observations $x_1, \ldots, x_T$ of $n \times 1$ random vector **x** with a common distribution $\mathcal{P} = \left\{ C'_\theta(x) \prod_{i=1}^T p_i(x_i) \mid \theta \in \Theta \right\}$ where $p_i(x_i)$ is marginal distribution associated with $x_i$, and the log-likelihood is

$$\mathcal{L}(\mathbf{W}, \theta) = \frac{1}{T} \log C'_\theta(\mathbf{x}) \prod_{i=1}^T p_i(x_i) = \frac{1}{T} \sum_{i=1}^T \log p_i(x_i) + \frac{1}{T} \log C'_\theta(\mathbf{x}). \qquad (15)$$

The representation is consist with two-phrase CCA framework in that the first term on the right of equation (15) implies mutual information of $x$ and that the second term is empirical estimation of entropy of $x$. It is not hard to proof that

$$\min D(p \| p_c) \Leftrightarrow \max \mathcal{L}(\mathbf{W}, \theta). \qquad (16)$$

5.3 Estimation of copula

Suppose the IC subspaces have been correctly determined by ICA and then we can identify the copula by minimizing the second term on the right of (9). Given a class of Copula $C(\mathbf{u}; \theta)$ with parameter vector $\theta \in \Theta$, and a set of sources $\mathbf{s} = (s_1, \ldots, s_n)$ identified from data set $X$, the problem is such a optimization one

$$\max_{\mathbf{W}, \theta} E_{p(\mathbf{s})} \left( C'(\mathbf{u}_\mathbf{s}; \mathbf{W}, \theta) \right). \qquad (17)$$

By using *Sklar* theorem, the copula to be identified has been separate with marginal distributions which are known except non-Gaussianity in ICA model. Therefore, the problem here is a semi-parametric one and only need identifying the copula.

Parametric method is adopted. First, we should select a hypothesis for copula among many choices of copulas available. The selection depends on many factors, such as priori knowledge, computational ease, and individual preference. Due to space limitations, only few of them are introduced here. For more detail please refer to [9].

When a set of sources **s** and a parametric copula is prepared, the optimization of (17) becomes an optimization problem which can be solved as follows:

$$\sum_{s_i=1}^{n} \frac{\partial C'}{\partial \theta}(\mathbf{u};\theta) = 0, \quad (20)$$

where many readily technique can be used.

## 6 Conclusions and further directions

In this paper, a framework named Copula Component Analysis for blind source separation is proposed as a generalization of ICA. It differs from ICA which assumes independence of sources that the underlying components may be dependent with certain structure which is represented by Copula. By incorporating dependency structure, much accurate estimation can be made, especially in the case where the assumption of independence is invalidated. A two phrase inference method is introduced for CCA which is based on the notion of multidimensional ICA. Many problems remain to be studied in the future, such as Identifiability of the method, selection of copula model and applications.